\documentclass[a4paper,11pt,twocolumn]{article}

\usepackage{amsmath}

\title{Comment on ``Partial conservation of seniority in semi-magic
  nuclei'' by Chong Qi}

\author{Kai Neerg{\aa}rd, Fjordtoften 17, 4700 N{\ae}stved, Denmark}

\date{}

\begin{document}

\maketitle

\noindent{\small
  \textbf{Abstract:} This review misrepresents parts of one of my
  publications and fails to mention important parts.
}

\medskip

Section~4.2 of the review~\cite{ref:Qi26} discusses at some length my
article~\cite{ref:Nee22}. Unfortunately it misrepresents parts of this
article and fails to mention important parts. My article deals with
the so-called partial seniority conserved or Escuderos-Zamick
multiplets, which are angular momentum multiplets with angular momenta
$I = 4$ and 6 in the state vector space of four fermions in an angular
momentum $j = \frac92$ shell found in numeric calculations by
Escuderos and Zamick to be stationary for every rotationally invariant
two-body interaction despite not being the only multiplets with these
angular momenta. They are defined by having seniority $v = 4$ and zero
coefficients of fractional parentage by the three-body multiplet with
$I = j$ and $v = 3$~\cite{ref:Esc06}. Their theory is also a major
theme in the review.

I define in my article a subspace
\begin{equation}\label{eq:Phi4}
  \Phi_4 = \text{span}_{m,m'} \, a_m^\dagger a_{m'} \Phi_0 .
\end{equation}
of the total four-body state vector space, where $\Phi_0$ is the
subspace of states with $I = 0$ and $a_{m}$ is the annihilator of a
particle in the state with magnetic quantum number $m$. The review
asserts that I prove that the Escuderos-Zamick multiplets are
orthogonal to $\Phi_4$ by showing that this space is invariant to
every rotationally invariant two-body interaction. This is a
misrepresentation of my analysis. That the Escuderos-Zamick multiplets
are orthogonal to $\Phi_4$ is shown in section II of my article to
follow directly from their definition. My computer algebraic
verification of invariance of $\Phi_4$ to every rotationally invariant
two-body interaction serves as a proof of the universal stationarity
of the Escuderos-Zamick multiplets that is alternative to computer
algebraic proofs based on interaction matrix elements between states
with $I = 4$ or 6~\cite{ref:Isa08,ref:Qi12,ref:Isa14}.

In the next paragraph the author of the review issues an
incomprehensible warning against my method and presents as his own
result what is established already in~\cite{ref:Nee22}, namely that
$\Phi_4$ is the complete orthogonal complement of the space
$\Phi_4^\perp$ spanned by the Escuderos-Zamick multiplets and the
$I = 10$ and 12 multiplets. A remark that he was unable to verify my
basic vectors for the $M = 0$ part $\Phi_{40}^\perp$ of $\Phi_4^\perp$
(called $[\Phi_1]$ in the review), where $M$ denotes the total
magnetic quantum number, suggests that they might not be correct. In
fact, by diagonalising the matrix $C$ given by my equation~(22), which
represents the restriction to $\Phi_{40}^\perp$ of the square of the
total angular momentum, and substituting in the resulting eigenvectors
the expressions in my equation~(7) for these basic vectors, one gets
exactly the coefficients listed in Table~4 of the review.

Despite being presented as a review, \cite{ref:Qi26} fails to mention
what might be the most remarkable result of my study, namely that in
the space $\Phi_4^\perp$ every rotationally invariant two-body
interaction acts as a linear combination of a constant and the square
of the total angular momentum. This entails the universal stationarity
of the Escuderos-Zamick multiplets and is thus a stronger result. It
gives rise to a relation between the energies of the Escuderos-Zamick
multiplets and the $I = 10$ and 12 multiplets that may be tested
experimentally; two of them determine the other two. And it implies
that there is a three-dimensional space of rotationally invariant
two-body interactions which vanish on every state in these four
angular momentum multiplets. (One of these interactions is the pairing
interaction, which vanishes due to the states' maximal seniority.)


\begin{thebibliography}{0}

\bibitem{ref:Qi26} C. Qi, arXiv:2602.11308, Eur. Phys. J. A
  \textbf{62:15} (2026).

\bibitem{ref:Nee22} K. Neerg{\aa}rd, Phys. Rev. C \textbf{106},
  024308 (2022).

\bibitem{ref:Esc06} A. Escuderos, L. Zamick, Phys. Rev. C \textbf{73},
  044302 (2006).

\bibitem{ref:Isa08} P. Van Isacker, S. Heinze, Phys. Rev. Lett.
  \textbf{100}, 052501 (2008).

\bibitem{ref:Qi12} C. Qi, Z. X. Xu, R. J. Liotta, Nucl. Phys. A
  \textbf{884-885}, 21 (2012).

\bibitem{ref:Isa14} P. Van Isacker, S. Heinze, Ann. Phys. (N. Y.)
  \textbf{349}, 73 (2014).

\end{thebibliography}
\end{document}